\documentclass[useAMS,referee,usenatbib]{biom}
\usepackage{graphicx}
\usepackage{amsmath}
\usepackage{booktabs}
\usepackage[figuresleft]{rotating}
\usepackage{multirow}
\setcitestyle{aysep={}} % no comma between author and year
\usepackage{graphicx}
\usepackage{color}
\usepackage{tabularx}
\usepackage{setspace}
\numberwithin{equation}{section}

\def\bSig\mathbf{\Sigma}

\title{Gene-based Association Analysis for Bivariate Time-to-event Data through Functional Regression with Copula Models}

\author{Yue Wei$^{1,+}$,
Yi Liu$^{2,+}$,
Wei Chen$^{3}$, and
Ying Ding$^{1,\ast}$\email{yingding@pitt.edu} \\
$^{+}$: Equal contribution \\
$^{\ast}$: Corresponding author \\
$^{1}$Department of Biostatistics, University of Pittsburgh, Pittsburgh, PA, U.S.A\\
$^{2}$Boehringer Ingelheim, Ridgefield, CT, U.S.A\\
$^{3}$Department of Pediatrics, Children's Hospital of Pittsburgh, Pittsburgh, PA, U.S.A \\
}

\begin{document}

%\date{{\it Received March} 2019. {\it Revised XXX} 2019.  {\it
%Accepted March} 2008.}

\pagerange{\pageref{firstpage}--\pageref{lastpage}}
\volume{xx}
\pubyear{xxxx}
\artmonth{xxxx}

\label{firstpage}

%  put the summary for your paper here

\begin{abstract}
Several gene-based association tests for time-to-event traits have been proposed recently, to detect whether a gene region (containing multiple variants), as a set, is associated with the survival outcome. However, for bivariate survival outcomes, to the best of our knowledge, there is no statistical method that can be directly applied for gene-based association analysis. Motivated by a genetic study to discover gene regions associated with the progression of a bilateral eye disease, Age-related Macular Degeneration (AMD), we implement a novel functional regression method under the copula framework. Specifically, the effects of variants within a gene region are modeled through a functional linear model, which then contributes to the marginal survival functions within the copula. Generalized score test and likelihood ratio test statistics are derived to test for the association between bivariate survival traits and the genetic region. Extensive simulation studies are conducted to evaluate the type-I error control and power performance of the proposed approach, with comparisons to several existing methods for a single survival trait, as well as the marginal Cox functional regression model using the robust sandwich estimator for bivariate survival traits. Finally, we apply our method to a large AMD study, the Age-related Eye Disease Study (AREDS), to identify gene regions that are associated with AMD progression. The method has been implemented and added into a newly developed R package \{CopulaCenR\}.
\end{abstract}


%  As usual, the \maketitle command creates the title and author/affiliations
%  display

\maketitle

%  If you are using the referee option, a new page, numbered page 1, will
%  start after the summary and keywords.  The page numbers thus count the
%  number of pages of your manuscript in the preferred submission style.
%  Remember, ``Normally, regular papers exceeding 25 pages and Reader Reaction
%  papers exceeding 12 pages in (the preferred style) will be returned to
%  the authors without review. The page limit includes acknowledgements,
%  references, and appendices, but not tables and figures. The page count does
%  not include the title page and abstract. A maximum of six (6) tables or
%  figures combined is often required.''

\section{Introduction}
\label{s:intro}

In genome-wide association studies (GWAS), single variant test is useful in detecting significant SNPs across the whole genome. Typically, after top loci are detected and confirmed in replication studies, regions around top variants will then be fine-mapped to further evaluate the disease loci. However, there are limitations of single variant tests. First, with each SNP  been tested individually, it may suffer from issues such as lack of power and multiple testing adjustment. Secondly, the true causal SNP may not be genotyped due to the technology or cost reason. Instead, a SNP that is close to the true causal variant is often captured. With partial linkage disequilibrium (LD), the observed effect size is likely to be smaller. Furthermore, most statistical approaches used for single variant test focus on common variants and can be too liberal for rare variants with low minor allele frequency (MAF). A threshold for MAF (e.g., $\ge 5\%$) is commonly applied when performing GWAS on single variants. To solve these problems, there has been increasing interests in developing gene-based tests in genetic association analysis, which can usually take the LD information within a region into account and are suitable for collapsing a set of variants with low MAF.

The statistical methods for gene-based association studies can be broadly classified as burden tests, kernel-based association tests and functional-regression-based methods. Burden test was first proposed by \citet{burden_test_Li} for binary traits to detect association with rare variants for common diseases. The general idea of burden test is based on collapsing rare variants in a genetic region to a summary variable, which is then used for testing the association with the phenotype. \citet{burden_test} extended the burden test to the censored time-to-event traits under Cox proportional hazards (PH) framework and \citet{burden_family} further developed burden test for survival traits in family-based designs. Traditional burden test suffers from the lack of power when heterogeneous genetic effects exist within a region, therefore several adaptive approaches have been proposed recently. For example, \citet{adaptive-burden} proposed to use mixed-model tests to make full use of genetic correlations across both samples and variants, and to gain power through ``data-driven'' weights which are adaptive to the direction of individual variant's effect. Another popular approach is the sequence kernel association test (SKAT), proposed by \citet{MichealWu_SKAT}. It is considered as a computationally efficient score test on variance component parameter to test for association between genetic variants in a region and different types of traits. In 2014, \citet{chen_SKAT} proposed the SKAT tests for survival traits in a Cox PH regression framework. One advantage of SKAT is that it can quickly calculate p-values by fitting the null model containing only the non-genetic covariates. A common limitation of the burden test and SKAT is the lack of effectively utilizing the LD information or in other words, correlations among genetic variants. \citet{Ruzong_quantative} first introduced the idea of functional regression (FR) for testing associations between genetic variants and quantitative traits. The FR-based model treats the effect of genetic variants as an unknown function of variants' actual physical positions in a functional linear model (FLM) and utilizes the LD information among close variants. It has been shown that for continuous, binary and censored traits, the FLM approach can be more powerful than SKAT or burden test in various scenarios \citep{Ruzong_quantative,Ruzong_binary,Ruzong_surv}.

This work is motivated by a genetic study for Age-related Macular Degeneration (AMD) to identify risk regions associated with AMD progression. AMD, a leading cause of blindness in the developed world \citep{AMD_prog_1,AMD_Prev}, is a bilateral and progressive neurodegenerative disease. Recently, multiple studies have been conducted to underpin the genetic causes for disease progression where the outcome of interest is the bivariate time-to-progression (e.g., time-to-late-AMD of the two affected eyes). For example, \citet{Sardell_AMD_CFH} and \citet{prediction_genetics} analyzed a small set of variants on AMD progression using a Cox PH model with the robust variance covariance estimate that takes the between-eye correlation into account. More recently, \citet{Yan2017_AMD} performed the first GWAS on AMD progression using a similar robust Cox PH model. Multiple risk loci have been identified to be associated with AMD progression. However, in all these studies, the focus is on common variants with MAF $\ge 5\%$. In this work, instead of evaluating the effect of single (common) variants on AMD progression, we are interested in identifying and assessing the gene regions associated with AMD progression. To our knowledge, there has been only one attempt so far on gene-based analysis for AMD progression, done by \citet{Ruzong_surv}. They applied the FR-based approach under the Cox PH model by using the left eye information only, due to the lack of methods for bivariate survival traits in gene-based analysis.

Bivariate or multivariate survival analysis has been studied for decades. Thorough reviews with examples can be obtained from \citet{Joe} and \citet{Hougaard_2000}. The Copula family is one of the earliest approaches for modeling bivariate survival data \citep{Clayton}. The idea is to model the joint distribution as a function of each marginal distribution together with a dependence parameter. Another popular approach for correlated survival data is the frailty model, of which the correlation is modeled through a common latent frailty variable \citep{Oaks}. The third approach is a marginal method \citep{LWL} such as the robust Cox model used in \citet{prediction_genetics} and \citet{Yan2017_AMD}, of which the variance-covariance matrix is estimated from a robust sandwich estimator.

In this work, we propose to develop and implement a novel gene-based association analysis method for bivariate survival traits based on functional regression under the copula framework. The paper is organized as follows. Section 2 introduces the proposed copula-based FLM method. Section 3 presents simulation studies for evaluating type-I error control and power performance under various settings. Section 4 demonstrates a real data analysis on AMD progression using the proposed method, and followed by the discussion of practical challenges and possible extensions in Section 5.

\section{Method} \label{s:Methods}
\subsection{Copula model for bivariate time-to-event data}
One of the earliest distribution families for modeling correlated bivariate measurements is the copula family \citep{Clayton}, originated from Sklar's Theorem \citep{sklar}, in which the joint distribution is modeled as a function of each marginal distribution together with a dependence parameter. Assume $U$ and $V$ are both uniformly distributed random variables, then a bivariate copula is a function defined as $\{C_\eta:  [0,1]^2 \rightarrow [0,1] : (u,v)\rightarrow C_\eta(u,v), \eta \in R\}$.  The parameter $\eta$ in the copula function describes the dependence between $U$ and $V$. By the Sklar's theorem \citep{sklar}, one can model the joint distribution by modeling the dependence parameter and the marginal distributions separately. This is the unique feature of copula functions that makes them attractive to use. The theorem is stated as: if marginal survival functions $S_{1}(t_1)=P(T_1 > t_1) $ and $S_{2}(t_2)=P(T_2 > t_2)$ for $T_1$ and $T_2$ are continuous, then there exists a unique copula function $C_\eta$ such that for all $t_1 \ge 0, t_2 \ge 0$, the joint survival function $S(t_1,t_2)$ can be written as
\begin{equation}
\label{copula}
S(t_1,t_2)= C_\eta((S_1(t_1),S_2(t_2)) , \ t_1, t_2 \geq 0.
\end{equation}
Define the density function for $C_\eta$ to be $c_\eta = \partial^2 C_\eta (u,v)/\partial \, u \partial \, v$, then the joint density function of $T_1$ and $T_2$ can be expressed as
\begin{align*}
f(t_1,t_2)=c_\eta(S_1(t_1),S_2(t_2))f_1(t_1)f_2(t_2) ,  \ t_1, t_2 \geq 0.
\end{align*}

\noindent
%The copula function is robust in modeling various dependence structures and has nice properties. For example, the rank-based dependence measurement Kendall's $\tau$ can be directly obtained as a function of $\eta$ in some copula models. %Tail dependence can also be determined through a copula function.

In this work, we choose to use the Archimedean copula family, which is one of the most popular copula families because of its flexibility and simplicity. For example, the rank-based dependence measurement Kendall's $\tau$ can be directly obtained as a function of $\eta$ in Archimedean copula models. Two most frequently used Archimedean copulas in survival analysis are: \\
{\em Clayton copula \citep{Clayton}}
\begin{eqnarray*}
\label{Clayton}
C_\eta(u,v)=(u^{-\eta}+v^{-\eta}-1)^{-1/\eta}, \ \eta \in (0,\infty),
\end{eqnarray*}
which models the lower tail dependence in survival functions; and \\
\noindent
{\em Gumbel-Hougaard copula \citep{Gumbel}}
\begin{eqnarray*}
\label{Gumbel}
C_\eta(u,v)=\exp\{-[(-\log u)^\eta+(-\log v)^\eta]^{1/\eta}\}, \ \eta \in [1,\infty),
\end{eqnarray*}
which models the upper tail dependence in survival functions.

We consider modeling the margins $S_1(\cdot)$ and $S_2(\cdot)$ inside the copula function using the Cox PH model, within which the effect of a gene region (captured by multiple single variants) is modeled through a FLM, as we describe below.

\subsection{Functional Linear Model Specification}
Assume $n$ individuals with $m$ variants being sequenced for a gene region. Physical positions for each variant within that region are denoted as $0 \leq u_1< \dots < u_m$ (which are typically standardized into $[0,1]$).  Let $G_i = (g_i(u_1), \dots, g_i(u_m))^\prime, \ g_i(u_j) \in (0,1,2), \ i=1,\dots,n, \ j=1,\dots,m,$ denote the genotype information for the $m$ variants of subject $i$, indicating the number of copies of the minor allele for each of these $m$ variants. Let $$(X_{1i},X_{2i}) = ((x_{1i1}, \dots, x_{1ip}), (x_{2i1}, \dots , x_{2ip})),$$ denote a $p \times 2$ matrix of bivariate covariates for subject $i$.
Then the hazard function for the $k$th margin ($k=1, 2$) under the PH-FLM can be written as
$$\lambda_{ki}(t)=\lambda_{k0}(t)\exp\left(X_{ki}^\prime \beta+ \int_{0}^{1} G_i(u)\gamma (u)du\right),$$
where $\lambda_{k0}(t)$ is the baseline hazard function for the $k$th margin, $\beta$ is a $p \times 1$ vector of coefficients for the non-genetic covariates, and $\gamma(u)$ is the genetic effect function of the genetic variant function $G_{i}(u)$ at position $u$. We assume both $\gamma(u)$ and $G_i(u)$ are smooth functions. Then the corresponding marginal survival functions are
\begin{align*}
S_{ki}(t) & =\exp(-\Lambda_{ki}(t))= \exp\left(-\int \lambda_{ki}(t)dt \right)\nonumber \\
&= \exp\left(-\int\lambda_{k0}(t)\exp\left(X_{ki}^\prime \beta+ \int_{0}^{1} G_i(u)\gamma (u)du \right)dt
\right), \ k=1,2.
\end{align*}

Next, we describe how we handle the two functions $G_i(u)$ and $\gamma(u)$, respectively.

\subsubsection{The genetic variant function (GVF) $G_i(u)$} If the genotype data are of good quality with low missing rate, we can simply use the observed genetic information to represent the GVF directly, such as $\hat{G}_i(u)=G_i = (g_i(u_1),\dots,g_i(u_m))$. In this case, the elements of $\hat{G}_i(u)$ take discrete values $0, 1, 2$.

If the genotype data have a fairly high missing rate or equivalent, the region has only a small number of variants being genotyped, one may consider applying an ordinary linear square smoother \citep{FDA_r} to obtain a continuous realization of $G_i$. Let $\phi(u)=(\phi_1(u),\dots,\phi_{B_G}(u))^\prime$ be a series of basis functions (e.g, B-spline or Fourier spline basis). Denote by $\Phi$ the $m\times B_G$ matrix with elements $\phi_b(u_j), \ b=1, \dots, B_G, \ j=1, \dots, m$. Then through a linear square smoother, an estimate of the GVF can be written as
\begin{eqnarray}
\label{GVF_est}
\hat{G}_i(u)=(g_i(u_1)),\dots,g_i(u_m))\Phi[\Phi^\prime\Phi]^{-1}\phi(u).
\end{eqnarray}
Under the situation with missing data, we have $$\hat{G}_i(u)=(g_i(u_1)),\dots,g_i(u_{m^\prime}))\tilde{\Phi}[\tilde{\Phi}^\prime\tilde{\Phi}]^{-1}\phi(u),$$ where $(g_i(u_1)),\dots,g_i(u_{m^\prime}))$ are the observed non-missing genotypes and $\tilde{\Phi}$ is the corresponding basis matrix evaluated at the non-missing genotypes.

\subsubsection{The genetic effect function (GEF) $\gamma(u)$}
The GEF $\gamma(u)$ is an unknown smooth function with an arbitrary form that we need to estimate. To do this, one can approximate it using a sieve approach with a linear combination of basis functions and coefficients. Define a series of $B_\gamma$ basis function by $\psi(u) = (\psi_1(u),\dots,\psi_{B_\gamma}(u))^\prime$ and a $B_\gamma \times 1$ vector $\gamma = (\gamma_1,\dots,\gamma_{B_\gamma})^\prime$, then $\gamma(u)$ can be approximated by
\begin{eqnarray}
\hat{\gamma}(u)=(\psi_1(u),\dots, \psi_{B_\gamma}(u))(\gamma_1,\dots,\gamma_{B_\gamma})^\prime.
\label{GEF_est}
\end{eqnarray}
To test whether the variants in a region, as a set, is associated with the outcome, it is to test whether the GEF is a zero function $\gamma(u)=0$, which is equivalent to test the null hypothesis: $H_0: \gamma_1=\dots=\gamma_{B_\gamma}=0$. Typically, the number of basis $B_\gamma$ is much smaller than the number of variants $m$ in the region being tested.

\subsubsection{The hazard function $\lambda(t)$ under the FLM}
Depending on whether or not to smooth the GVF $G(u)$, we propose two types of functional regression models for the hazard function. The first option is to smooth both $G(u)$ and $\gamma(u)$. By replacing $G(u)$ and $\gamma(u)$ with their approximated values in \eqref{GVF_est} and \eqref{GEF_est}, the hazard function for each margin can be written as
\begin{eqnarray}
\label{lambda_1}
\lambda_{ki}(t) &  =  &\lambda_{k0}(t) \exp \left(X_{ki}^\prime \beta + (g_i(u_1),\dots,g_i(u_m)) \Phi[\Phi^\prime\Phi]^{-1}\gamma  \int_0^1 \phi(u)\psi(u) du \right) \\
& = & \lambda_{k0}(t)\exp(X_{i}^\prime\beta+M_i^\prime \gamma), \nonumber
\end{eqnarray}
where $M_i^\prime = (g_i(u_1),\dots,g_i(u_m)) \Phi[\Phi^\prime\Phi]^{-1} \int_0^1 \phi(u)\psi(u) du$. The integral $ \int_0^1 \phi(u)\psi(u) du$ can be readily calculated using the R package \{fda\} \citep{FDA_r} once the basis functions are chosen.

Another option is to smooth $\gamma(u)$ only. In this case, we directly replace $G_i(u)$ by the observed $G_i=(g_i(u_1),...,g_i(u_m))^\prime$ and replace $\gamma(u)$ by \eqref{GEF_est}, which yields
\begin{eqnarray}
\label{lambda_2}
\lambda_{ki}(t) & = & \lambda_{k0}(t)\exp \left(X_{ki}^\prime \beta + \left[\sum_{j=1}^{m}(g_i(u_j) \times (\psi_1(u_j),\dots,\psi_{B_\gamma}(u_j))\right] (\gamma_1,\dots,\gamma_{B_\gamma})^\prime \right) \\
& = & \lambda_{k0}(t)\exp(X_{i}^\prime\beta+M_i^\prime \gamma), \nonumber
\label{beta_only}
\end{eqnarray}
where $M_i^\prime=\sum_{j=1}^m g_i(u_j) \times (\psi_1(u_j),\dots,\psi_{B_\gamma}(u_j))$ is a fully observed term.

Two frequently used basis functions are B-splines and Fourier splines. The B-spline basis \citep{deboor_spline} is a series of non-periodic functions with polynomial segments joint at values called knots. The segments have specifiable smoothness across every breaks. Advantages of applying the B-spline basis include its fast computation and great flexibility in structure. The Fourier splines consist of a set of periodic functions, with the basis function being $\Phi_1(u)=1$, $\Phi_{2r}(u)=\cos(2\pi ru)$ and $\Phi_{2r+1}(u)=\sin(2\pi ru)$ for a pre-specified $r$ \citep{deboor_spline}.

\subsection{Bivariate FLM under the copula framework}
%In this work, the obtained genotype data from AREDS have passed strict quality control procedures and with no missings. Thus we do not smooth the GVF and use the observed $G_i$ value for $G_i(u)$. So we only smooth the GEF. Note that, instead of estimating $m$ parameters for $m$ variants, we have only $B_\gamma$ ($\ll$ m in most cases) parameters to estimate.\\

Let $(T_{1i},T_{2i})$ and $(C_{1i},C_{2i})$ denote the bivariate survival times and censoring times for subject $i$, respectively. Denote by $\Delta_i=(\Delta_{1i},\Delta_{2i})$ the corresponding censoring indicator. We consider right censoring and assume that given covariates, $(T_1,T_2)$ and $(C_1,C_2)$ are independent. Then for each subject, we observe
\begin{align*}
D_i=\{(Y_{1i},Y_{2i}, \Delta_{1i},\Delta_{2i},X_{1i}, X_{2i}, G_i): Y_{ki}=\min(T_{ki},C_{ki}), \Delta_{ki}=I(T_{ki}\le C_{ki}), k=1,2\}.
\end{align*}
Denote by $\theta$ all the parameters in $S(t_{1},t_{2})$, then the joint likelihood for the observed data $\{D_i\}_{i=1}^n$ can be written as
\begin{equation} \label{joint_likelihood}
\begin{split}
 L(\theta ; D  &  =(Y_{1},Y_{2}, \Delta_{1},\Delta_{2},X_{1}, X_{2}, G)) \\
 &  = \prod_{i=1}^n \, f(y_{1i},y_{2i})^{\delta_{1i}\delta_{2i}}  \times\left[-\frac{\partial S(y_{1i},y_{2i})}{\partial y_{1i}}\right]^{\delta_{1i}(1-\delta_{2i})} \\
& \times \left[-\frac{\partial S(y_{1i},y_{2i})}{\partial y_{2i}}\right]^{(1-\delta_{1i})\delta_{2i}} \times S(y_{1i},y_{2i})^{(1-\delta_{1i})(1-\delta_{2i})},
\end{split}
\end{equation}
where $(\delta_{1i},\delta_{2i}) \in \{(0,0),(0,1),(1,0),(1,1)\}$. Then under the copula framework (\ref{copula}), with a FLM for the genetic effect, the joint survival function can be further written as
\begin{equation*}
S(y_{1i},y_{2i}) = C_\eta (\exp\left(-\int_0^{y_{1i}}\lambda_{1i}(t)dt\right),\exp\left(-\int_0^{y_{2i}}\lambda_{2i}(t) dt\right) ),
\end{equation*}
where $\lambda_{ki}(t)$ can be modeled by either (\ref{lambda_1}) or (\ref{lambda_2}), depending whether to smooth the GVF $G_i(u)$ or not.

\subsubsection{Generalized score test}
We are interested in testing whether a gene region is associated with the disease progression, after adjusting for other risk factors. To accomplish this, we separate the entire parameter $\theta$ into two parts: $\theta_1$, the parameter of interest; and $\theta_2$, the nuisance parameter. Under the bivariate FLM-based copula model, we have $\theta_1=\gamma$, which is the gene/region effect, and $\theta_2=(\beta, \eta, \lambda_{10}, \lambda_{20})$ contains the rest of parameters in the likelihood. Then the null hypothesis can be formulated as
\begin{align}
\label{h0_null}
H_0 : \theta_{1}=(\gamma_1,\dots,\gamma_{B_\gamma})^\prime = 0 \mbox{ and } \theta_2 \mbox{ is arbitrary}.
\end{align}

In GWAS, score test is typically preferred than other likelihood-based tests due to its computational advantage. This is because only one null model (without any SNP) needs to be fitted for the score test. We now describe the generalized score test under the FLM-based copula framework for testing (\ref{h0_null}).

Assume $\hat{\theta}_{0}=(\theta_{1}=0, \theta_{2}=\hat{\theta}_{20})$ is the restricted maximum likelihood estimate (MLE) of $\theta$ under the restriction of $\theta_{1}=0$, solved from the joint likelihood under (\ref{joint_likelihood}), then the corresponding score function is
\begin{align*}
U(\hat{\theta}_0) = \frac{\partial}{\partial \theta}\, \log L(\theta; D)\biggr\rvert_{\theta=\hat{\theta}_0} = (U_1^\prime(\hat{\theta}_0), U_2^\prime(\hat{\theta}_0))^\prime=(U_1^\prime(\hat{\theta}_0),0^\prime)^\prime ,
\end{align*}
where $U_l(\cdot) ={\partial \log L}/{\partial \theta_l}, \ l=1, 2$, and the Fisher's information is
\begin{align*}
\mathcal{I}(\hat{\theta}_0)=-E\left[\frac{\partial ^2}{\partial \theta^T \partial \theta} \, \log  L(\theta;D)\right]\biggr\rvert_{\theta=\hat{\theta}_0} =
 \begin{bmatrix}
    \mathcal{I}_{11}(\hat{\theta}_0)&\mathcal{I}_{12}(\hat{\theta}_0)\\
    \mathcal{I}_{21}(\hat{\theta}_0)& \mathcal{I}_{22}(\hat{\theta}_0)
    \end{bmatrix},
\end{align*}
with $\mathcal{I}_{11}, \mathcal{I}_{12}, \mathcal{I}_{21}$ and $\mathcal{I}_{22}$ being partitions of the information matrix $\mathcal{I}$ by $\theta_1$ and $\theta_2$.

Finally, by using the observed information matrix $\mathcal{J}(\hat{\theta}_0)$ ($\mathcal{J}(\theta)=-\frac{\partial^2 \log L(\theta;D)}{\partial \theta^\prime \partial \theta}$) to approximate $\mathcal{I}(\hat{\theta}_0)$, the generalized score statistic for testing (\ref{h0_null}) can be constructed as follows
\begin{align*}
Q_s  = U_1^\prime(\hat{\theta}_0)\mathcal{J}^{11}(\hat{\theta}_0)U_1(\hat{\theta}_0),
\end{align*}
where $U_1^\prime(\hat{\theta}_0)$ is a $B_\gamma \times 1$ vector and $\mathcal{J}^{11}=(\mathcal{J}^{-1})_{11}=(\mathcal{J}_{11}-\mathcal{J}_{12}\mathcal{J}_{22}^{-1}\mathcal{J}_{21})^{-1}$ is a $B_\gamma \times B_\gamma$ matrix.

We can use similar numerical approximation techniques as proposed in \citet{CopulaRC}, such as using the Richardson's extrapolation to approximate the score function and observed information matrix. The score test statistic asymptotically follows a $\chi^2$ distribution with $B_\gamma$ degrees of freedom under the null.

\subsubsection{Likelihood ratio test}
Different from the GWAS case, where the computational efficiency is a key factor for deciding the test procedure,  we have a lot less number of tests in the gene-based test situation ($\sim 20K$ genes vs millions of variants in GWAS). Therefore, an alternative approach to the generalized score test is to perform the likelihood ratio test (LRT). Specifically, the LRT statistic can be written as
\begin{equation*}
Q_l = -2 (\log L(\hat{\theta}_0) - \log L(\hat{\theta})),
\end{equation*}
where $L(\hat{\theta})$ is the the unrestricted maximum likelihood value of the joint likelihood and  $L(\hat{\theta}_0)$ is the restricted maximum likelihood under $H_0$. $Q_l$ also asymptotically follows a $\chi^2$ distribution with degrees of freedom $B_\gamma$.

\section{Simulation Study}
\label{s:Simu}
In this section, we performed simulation studies to evaluate the performance of the proposed method in terms of type-I error control and power. Two scenarios were considered: (1) a mixture of common (MAF$> 5\%$) and rare causal variants (MAF$\in [1\%, 5\%]$); (2) all causal variants are rare (MAF $\in [1\%, 5\%]$). We reported the results of copula-based functional regression model using score test (Cop-Score) and LRT (Cop-LRT), and compared with the following methods: (a) FLM under Cox PH assuming the bivariate traits are independent (Cox-Ind), or (b) using the ``robust sandwich estimator'' for the variance-covariance matrix (Cox-Rst), (c) univariate FLM under Cox PH (Cox-FLM), (d) Burden test, and (e) SKAT test. The last two methods are from R package \{seqMeta\} \citep{seqMeta}. To evaluate the performance of univariate methods (c)-(e), only one margin (left eye) from simulated data was used.

\subsection{Data generation}
In our simulation, 100 genotype datasets, each with a sample size 1000 were simulated. For each genotype set, 1000 phenotype datasets of bivariate survival times and one non-genetic covariate were generated, which lead to a total of 100,000 datasets. Similar simulation approach was used in \citet{chen_SKAT}, which has been shown to produce valid comparisons between methods.

Genetic data were generated from European ancestry of 10,000 haplotypes covering 1Mb regions, simulated by Yun Li at the University of North Carolina \citep{Ruzong_surv}. Calibrated coalescent model as programmed in COSI was used to generate the haplotypes with linkage disequilibrium (LD) information \citep{schaffner_LD}. With 10,000 haplotypes, we chose a genetic region of length 6 Kb and 30 Kb for all and rare only variants scenarios, respectively. For both scenarios, the regions contain around 20 variants. A random mating technical was then applied to generate genetic information for 1000 subjects.

Bivariate time-to-event phenotypes were generated from a Clayton Weibull model as follows. Recall that under a copula model $S(t_1,t_2)=C_\eta(S_1(t_1),S_2(t_2))$, $U=S_1(T_1)$, $V=S_2(T_2)$ each follows a uniform distribution $U[0,1]$. Define $W_v(u) =h(u,v) = P(U \leq u |V=v) = \partial C_{\eta}(u,v)/\partial v$. First, we generated $v_i$ and $w_i$ from two independent standard uniform distributions. Then let $w_i=h(u_i,v_i)(=C_{\eta}(u_i,v_i)/\partial v_i$) and solve for $u_i$ from the inverse function $h^{-1}$. Finally, we obtained two survival times $t_{1i}$ and $t_{2i}$ from $S^{-1}_1(u_i)$ and $S^{-1}_2(v_i)$ respectively. The scale and shape parameters in the baseline Weibull distribution were set to be $\lambda = 0.1$ and $k=2$, same for both margins. For Clayton copula, the dependence strength is determined by Kendall's $\tau$. Here we chose $\tau=$ 0.05, 0.4 and 0.8 to represent weak to strong dependence between two marginal survivals. Censoring times $c_{1i}$ and $c_{2i}$ were generated from uniform distribution $U(0,C)$ with $C$ chosen to yield a $50\%$ censoring rate. The non-genetic covariate $X_{ng,k}(k=1,2)$ was generated from $N(6,2^2)$.

For type-I error control simulations, we assumed there is neither a genetic effect ($\gamma(u) = 0$) nor a non-genetic effect ($\beta = 0$). The type-I error was evaluated at various $\alpha$ levels: $0.05, 0.01, 10^{-3}$ and $10^{-4}$, respectively. For power analysis, we generated data to evaluate both homogeneous genetic effects (genes with effects in the same direction) and heterogeneous genetic effects (genes with effects in opposite directions). Similar as in \citep{Ruzong_surv}, the effect size for each causal variant (i.e., $\gamma(u)$) was chosen to be a constant that depends on MAF: $c\frac{|log_{10}MAF|}{2}$, where $c=0.4,0.3,0.25$ for scenarios with $10\%$, $20\%$ and $30\%$ of causal variants in a given region, respectively.

\subsection{Type-I error}
We chose B-splines with 5 basis for all FLM methods. Table \ref{tab:type1_FR} presents the type-I error results for the scenario with both common and rare variants. As expected, the independent Cox FLM (Cox-Ind) yields inflated type-I errors and the inflation becomes more severe when data are more dependent. With robust variance-covariance estimates, Cox-Rst still shows some inflation in type-I error, especially under small $\alpha$ tail. Similar observations were found in testing single variants with bivariate survival outcomes \citep{CopulaRC}. Within the copula-based FLM, both the score test (Cop-Score) and LRT (Cop-LRT) produce satisfactory-controlled type- I errors.

\begin{table}
\caption {Type-I error at various dependence levels from Clayton copula with Weibull margins for testing gene regions that contain both common and rare variants.}
\label{tab:type1_FR}
\resizebox{\textwidth}{!}{% Resize table to fit within \linewidth horizontally
  \begin{tabular}{llllllllll}
\toprule
  &  &
      \multicolumn{4}{c}{Bivariate FLM} & &
      \multicolumn{3}{c}{Univariate} \\
      \cline{3-6}
      \cline{8-10}
	 \textbf{$\tau$}	  & $\alpha$ level  & Cop-Score & Cop-LRT & Cox-Ind &  Cox-Rst  & & CoxFLM & Burden & SKAT\\
	  \cline{1-10}
	  \multirow{4}{*}{$0.05$}
				 &0.05      & 0.0547 & 0.0524 & 0.0673 & 0.0626 & & 0.0523  & 0.0492 & 0.0494\\
				 &0.01    & 0.0114 & 0.0102 & 0.0152 & 0.0150 & & 0.0104  & 0.0105 & 0.0099\\
				&0.001  & 0.0014 & 0.0012 & 0.0020 & 0.0022 & & 0.0012 & 0.0010& 0.0012 \\
				&0.0001 & 0.00015 & 0.00010 & 0.00033 & 0.00047 & & 0.00018 & 0.00010 & 0.00009\\
	\cline{2-10}
	  \multirow{4}{*}{$0.4$}
				 &0.05      & 0.0527 & 0.0522 & 0.2119 & 0.0653 & & 0.0515 & 0.0495& 0.0504 \\
				 &0.01    &0.0105 & 0.0103 & 0.0832 & 0.0162 & & 0.0104 & 0.0099& 0.0099 \\
				&0.001  & 0.0011 & 0.0011 & 0.0214 & 0.0025 & & 0.0012 & 0.0010& 0.0011 \\
				&0.0001 & 0.00015 & 0.00012 & 0.00521 & 0.00047 & & 0.00010 & 0.00012& 0.00009 \\
	\cline{2-10}
	  \multirow{4}{*}{$0.8$}
				 &0.05      & 0.0534 & 0.0521 & 0.3293 & 0.0628 & & 0.0526 &  0.0500& 0.0511\\
				 &0.01    & 0.0110 & 0.0105 & 0.1644 & 0.0159 & & 0.0107 & 0.0097 & 0.0101 \\
				&0.001  & 0.0011 & 0.0011 & 0.0594 & 0.0021 & & 0.0010 & 0.0011 & 0.0012\\
				&0.0001 & 0.00012 & 0.00014 & 0.02103 & 0.00042 & & 0.00010 & 0.00012 & 0.00013\\
\bottomrule

 \end{tabular}}
\end{table}

Table \ref{tab:type1_FR_rare} presents the type-I error results for testing regions that contain only rare variants (MAF $\in [1\%, 5\%]$). The overall performance is similar to the all variants scenario. The inflation in the robust Cox model (Cox-Rst) is slightly more compared to the all variants scenario, which is also consistent with the observation in \citet{CopulaRC}.

\begin{table}
\caption{\label{tab:type1_FR_rare}Type-I error at various association levels from the Clayton copula with Weibull margins for testing gene regions that contain only rare variants.}
\resizebox{\textwidth}{!}{% Resize table to fit within \linewidth horizontally
  \begin{tabular}{llllllllll}
\toprule
  &  &
      \multicolumn{4}{c}{Bivariate FLM} & &
      \multicolumn{3}{c}{Univariate} \\
      \cline{3-6}
	  \cline{8-10}
	 \textbf{$\tau$}	  & $\alpha$ level  & Cop-Score & Cop-LRT & Cox-Ind &  Cox-Rst  & & CoxFLM & Burden & SKAT\\
	  \cline{1-10}
	  \multirow{4}{*}{$0.05$}
				 &0.05      & 0.0539 & 0.0540 & 0.0690 & 0.0677 & & 0.0516 & 0.0500 & 0.0490\\
				 &0.01    & 0.0115 & 0.0118 & 0.0160 & 0.0175 & & 0.0105 &0.0099 & 0.0094\\
				&0.001  & 0.0011 & 0.0011 & 0.0021 & 0.0035 & & 0.0010 & 0.0010 & 0.0009\\
				&0.0001 & 0.00014 & 0.00011 & 0.00030 & 0.00135 & & 0.00010 &0.00009 & 0.00010\\
	\cline{2-10}
	  \multirow{4}{*}{$0.4$}
				 &0.05      & 0.0541 & 0.0537 & 0.2151 & 0.0674 & & 0.0521 & 0.0505 & 0.0499\\
				 &0.01    &0.0113 & 0.0110 & 0.0855 & 0.0173 & & 0.0110 & 0.0101 & 0.0099\\
				&0.001  & 0.0013 & 0.0012 & 0.0226 & 0.0032 & & 0.0011 & 0.0011 & 0.0008 \\
				&0.0001 & 0.00012 & 0.00012 & 0.00628 & 0.00063 & & 0.00010 & 0.00011 & 0.00014\\
	\cline{2-10}
	  \multirow{4}{*}{$0.8$}
				 &0.05      & 0.0534 & 0.0517 & 0.3305 & 0.0646 & & 0.0519 & 0.0502 & 0.0502\\
				 &0.01    & 0.0110 & 0.0101 & 0.1671 & 0.0151 & & 0.0112 &0.0104 & 0.0103\\
				&0.001  & 0.0011 & 0.0010 & 0.0609 & 0.0022 & & 0.0012 &0.0010 & 0.00091\\
				&0.0001 & 0.00012 & 0.00006 & 0.02122 & 0.00035 & & 0.0001 & 0.00008 & 0.00010\\
\bottomrule
 \end{tabular}}
\end{table}

\subsection{Empirical power}

\begin{figure}[!ht]
\centering
\includegraphics[scale=0.75]{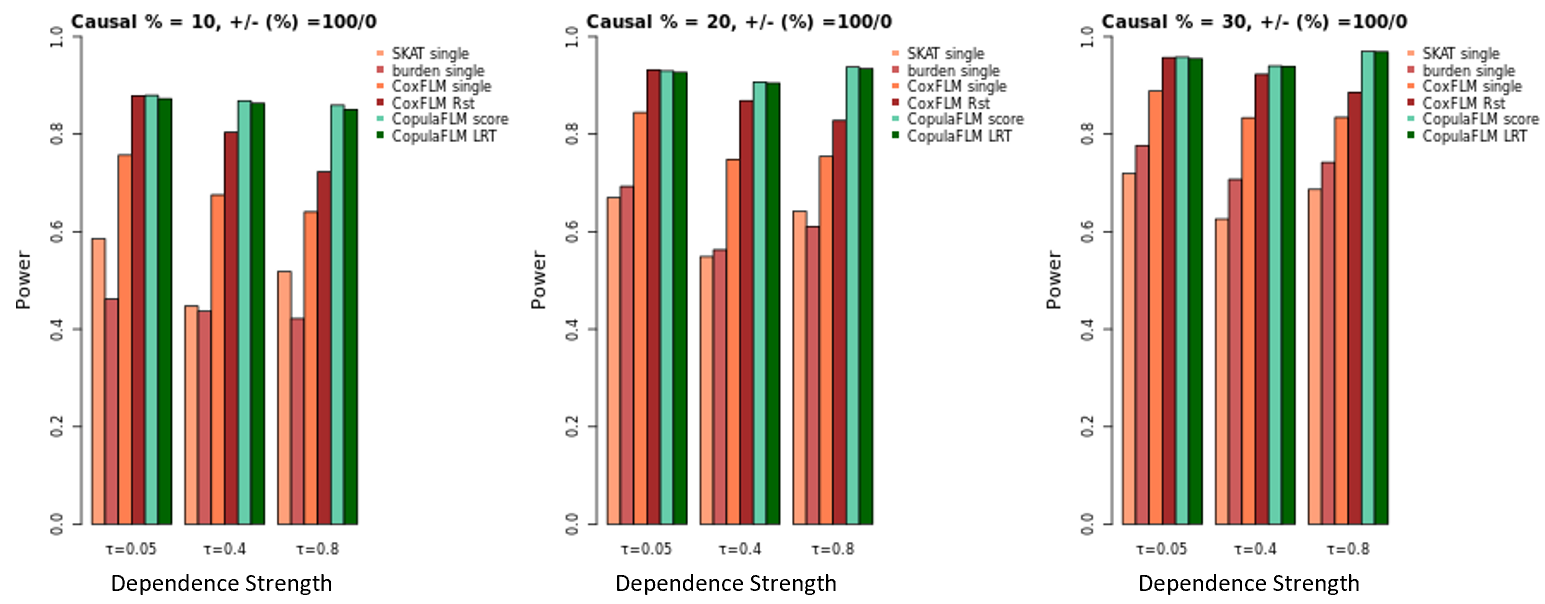}\\
\includegraphics[scale=0.75]{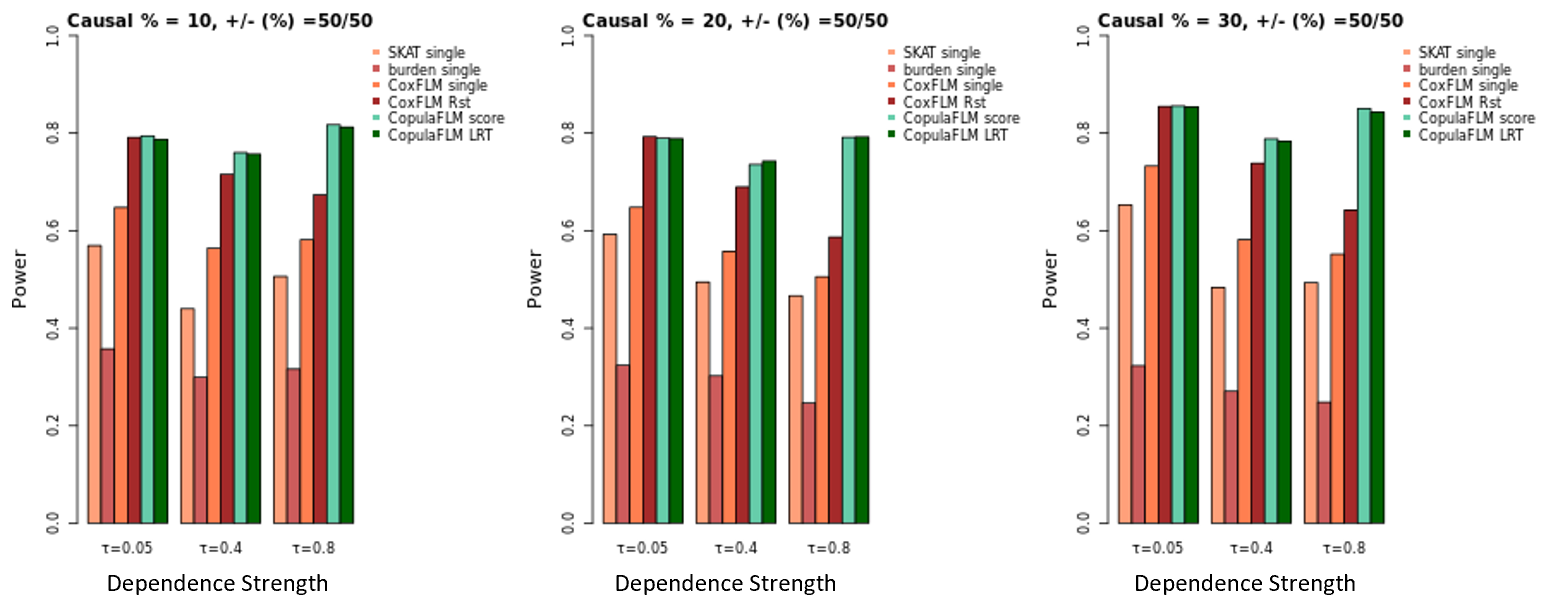}
\caption{\label{power_FR}Empirical power analysis for 1000 gene regions that contain both common and rare variants at various dependence strength levels}
\end{figure}

Figure \ref{power_FR} illustrates the power bars over different dependence levels and genetic effect sizes for the mixture of common and rare variants scenario. Given the independent Cox model cannot control type-I error at all, we did not include it in the power analysis. Overall, the bivariate approaches produce higher power than any univariate approach. When the two margins are highly dependent, the copula-based tests achieve higher power than the Cox-Rst method. In general, the score and LRT tests provide similar results, which is as expected. We also notice that, when there are heterogeneous genetic effects in a region (the bottom panel), the power of burden test drops significantly.

\begin{figure}[!ht]
\centering
\includegraphics[scale=0.8]{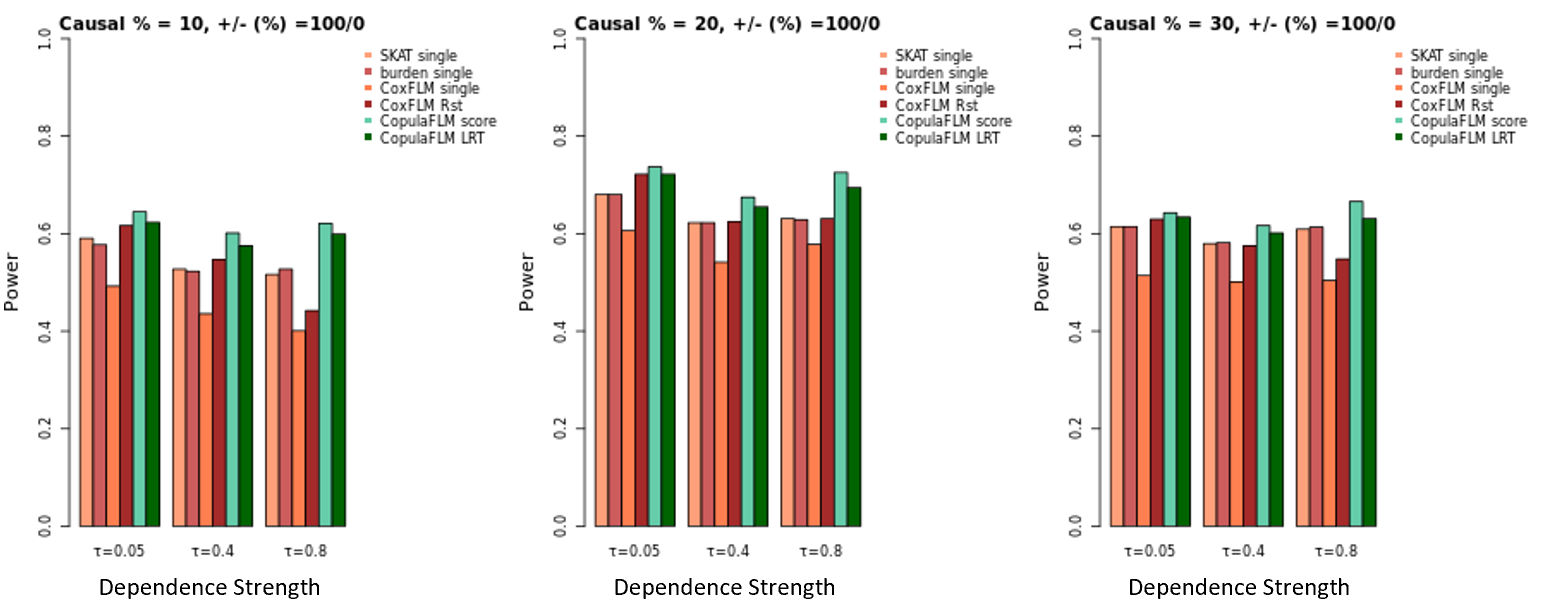}\\
\includegraphics[scale=0.8]{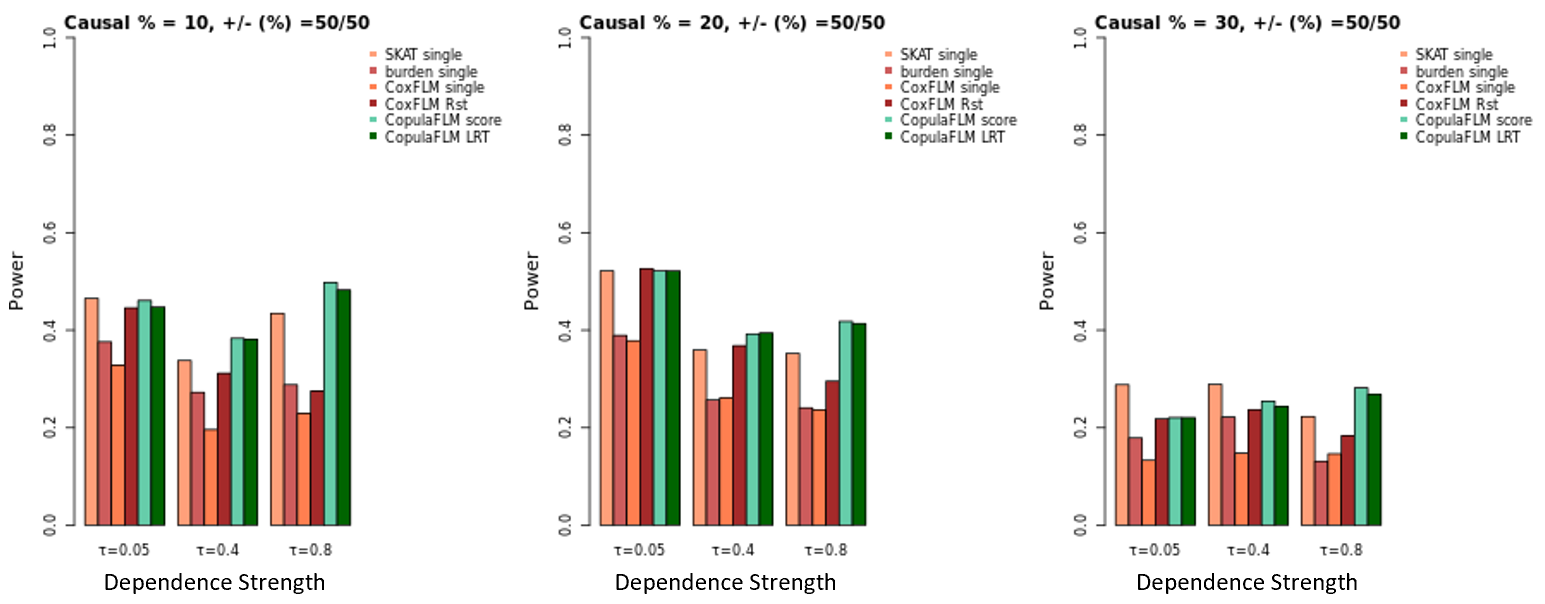}
\caption{\label{power_FR_rare}Empirical power analysis for 1000 gene regions that contain only rare variants at various dependence strength levels}
\end{figure}

Figure \ref{power_FR_rare} shows the power analysis for the rare variants only scenario. It can be seen that when only rare variants are considered, the univariate Cox FLM model (CoxFLM) does not perform well as compared to other univariate methods, which is different from the case when all variants are included. In contrast, SKAT maintains relatively high power for both cases of homogeneous and heterogeneous genetic effects. Burden test shows comparable results when the genetic effects are homogeneous, and the power drops when the effects are heterogeneous. As for the bivariate methods, Cox-Rst model performs well when the dependence between two margins is weak, but as the dependence increases, the power tends to decrease. Overall, our copula-based bivariate methods achieve high power in all the scenarios.

\section{Application on AREDS data}
\label{s:Analysis}
\subsection{Examination on the Reported Four Regions from Previous Publications}
AREDS is a major clinical trial sponsored by the National Eye Institute to study the risk factors for AMD progression \citep{AREDS}. The bilateral nature makes it a perfect example for the demonstration of our proposed methods. Over the past few years, many case-control GWAS studies have identified multiple SNPs associated with AMD susceptibility \citep{Wei_PNAS_2010,AMD_genetic_2013, AMD_genetic_2016} and recent genetic studies on AMD progression identified SNPs that are associated with disease progression \citep{prediction_genetics, Yan2017_AMD}. Table \ref{tab:AREDS_genes} presents top four genetic regions which contain SNPs associated with AMD risk identified by the traditional single-marker GWAS in \citet{AMD_genetic_2016}. These four regions were also reported to be top gene regions containing variants associated with AMD progression in \citet{Yan2017_AMD}. Therefore, we first examined these four candidate regions specifically. Location of gene regions were extracted based on GRCh37/hg19 assembly from UCSC Genome Browser. Variants within $+/-5 Kb$ of the region boundary were included in the analysis.

We included all Caucasian participants with neither eye progressed at the time of enrollment into the study. For bivariate approaches, time-to-progression was calculated for each eye of each patient. For univariate approaches, only left-eye data were used.  A total of 2296 subjects were included in the analysis. The baseline age and disease severity score (on a continuous scale ranging from 1 to 8) were included in the regression part of the model as the non-genetic risk factors.

\begin{table}
\caption{Four candidate gene regions from published single variant case-control GWAS AMD studies}
\label{tab:AREDS_genes}
\begin{tabular}{lllll}
\hline
Region & Chr & StartPos** & EndPos** & Number of SNPs \\
\hline
\textit{CFH} & 1 & 196,621,008 & 196,716,634 & 281 \\
\textit{C2-CFB-SKIV2L} & 6 & 31,65,562 & 31,937,532 & 250 \\
\textit{ARMS2} regions* & 10 & 124,134,094 & 124,274,424 & 541 \\
\textit{C3} & 19 & 6,677,715 & 6,730,573 & 294\\

\hline
 \multicolumn{2}{l}{\small * \textit{PLEKHA1,MIR3941,ARMS2,HTRA1}}\\
 \multicolumn{4}{l}{{\small ** Actural regions are selected with reference start and end pos $+/- 5K$ (hg19)}}\\
\end{tabular}
\end{table}

We fitted a Clayton copula with Weibull margins to perform the gene-based analysis. B-spline was used to smooth the genetic effect function. The number of basis is usually decided based upon the sample size (or equivalently, the total number of events in survival analysis) or through cross validations. In our analysis, we examined three different number of bases: 5, 6 and 7. A mixture of common (MAF$> 5\%$) and rare causal variants (MAF$\in [1\%, 5\%]$) were tested. Similar to the simulation studies, we compared our proposed copula-based FLM score test to the Cox robust model, the univariate Cox FLM, SKAT and burden test. Since the copula-based FLM LRT has shown virtually identical results to the score test in the simulations, we did not include it here.

Table \ref{tab:4_gene_flm} presents \textit{p}-values of testing the four candidate gene regions that contain both common and rare variants. The bivariate approaches, i.e., copula-based FLM (Cop-Score) and robust Cox (Cox-Rst) identify \textit{CFH}, \textit{C2-CFB-SKIV2L} and \textit{ARMS2} regions at the significance level of 0.05, with smaller \textit{p}-values produced from Cop-Score. For the \textit{C3} region, both bivariate methods have shown marginal significance. Due to ignoring data from the right eye, none of the univariate approaches identifies all four gene regions, with the best case scenario, SKAT, identifying three gene regions except the \textit{ARMS2} region. The univariate Cox-FLM identifies two gene regions: \textit{CFH} and \textit{ARMS2}, while the burden test only shows that \textit{CFH} is significant. Note that for the three approaches that involve FLM, varying the number of basis from 5 to 7 produces similar results in this data set.

\begin{figure}[!ht]
\centering
\includegraphics[scale=0.7]{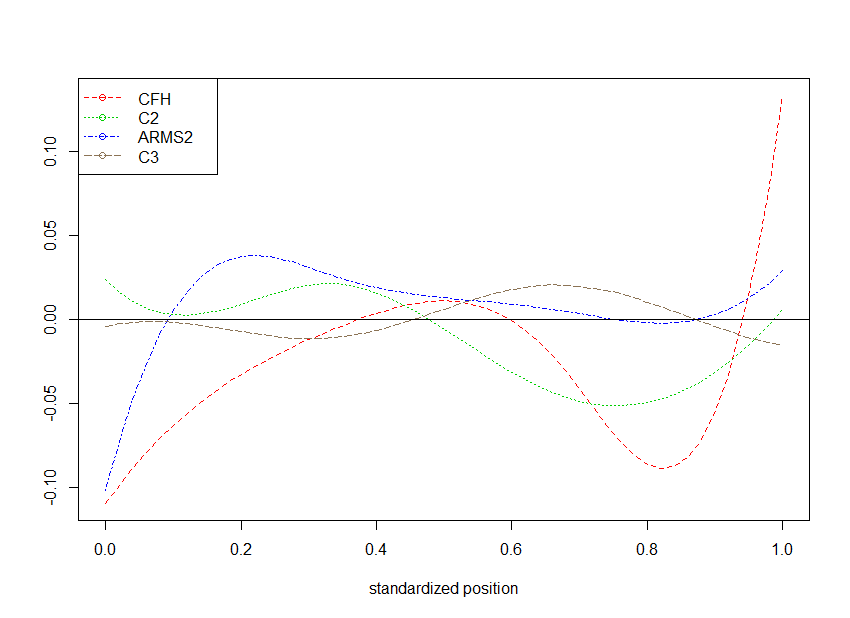}
\caption{\label{gamma}Estimated genetic effect functions for four candidate gene regions from the copula-FLM approach with six number of B-spline basis}
\end{figure}

Figure \ref{gamma} presents the estimated genetic effect functions for the four gene regions under the six B-spline bases scenario. The \textit{CFH} shows the largest effect size with most of the positions having a negative effect, and it explains the smallest p-values obtained from the FLM based methods for this region. Since most of the effects seem to be homogeneous in the \textit{CFH} region, the burden test maintains the power by providing a p-value as small as $7.11 \times 10^{-4}$. For other regions, they all show heterogeneous effects with some positions having positive effects while others having negative effects, so the burden test loses power under these situations. The \textit{ARMS2} and \text{C2} regions both show reasonable amount of effects and thus the copula-based FLM method has detected these two regions to be significant. The flat curve presented by the \textit{C3} region also explains the marginal significance from the copula-FLM approach.

\begin{table}
\caption{Results from different gene-based tests on four candidate regions (including both common and rare variants) using the AREDS data}
\label{tab:4_gene_flm}
\resizebox{\linewidth}{!}{% Resize table to fit within \linewidth horizontally
  \begin{tabular}{lllllllll}
\toprule
  &  &
      \multicolumn{4}{c}{Bivariate FLM} & &
      \multicolumn{1}{c}{Univariate} \\
    \cline{4-5}
    \cline{7-9}
	Gene & Chr & basis \#  & Cop-Score & Cox-Rst & & CoxFLM & Burden  & SKAT \\
	\cline{1-9}
  &  & 5 & $6.79 \times 10^{-10}$ &  $1.05 \times 10^{-7}$   &   & $3.82 \times 10^{-4}$\\
\textit{CFH}  & 1   & 6  & $9.93 \times 10^{-10}$ & $1.43 \times 10^{-7}$ & & $4.14 \times 10^{-4}$ & $7.11 \times 10^{-4}$ & 0.01 \\
              &     & 7 & $3.24 \times 10^{-9}$ &  $3.56 \times 10^{-7}$ & & $8.14 \times 10^{-4}$\\
	
	\cline{1-9}
	& & 5  & $ 2.02\times 10^{-3}$ & $8.17 \times 10^{-3}$ &  & 0.13 \\
	\textit{C2-CFB-SKIV2L} & 6 & 6  & $3.74 \times 10^{-3}$ & 0.01 & & 0.19 & 0.43 & 0.02 \\
	& & 7 & $ 6.81 \times 10^{-3}$ &  0.02 & & 0.23\\
	\cline{1-9}
	& & 5  & $5.40 \times 10^{-4}$ & 0.01 & & 0.03\\
	\textit{ARMS2} regions &10 & 6 & $8.06 \times 10^{-4}$ & 0.01 & & 0.03 & 0.88 & 0.19 \\
	& & 7 & $4.81 \times 10^{-5}$ & $5.91 \times 10^{-3}$ & & 0.01\\
	\cline{1-9}
	& & 5  & 0.05 & 0.03 & & 0.14\\
\textit{C3} &19 & 6 & 0.05 & 0.04 & & 0.15 & 0.77& $1.37 \times 10^{-3}$  \\
    & & 7 & 0.10 & 0.06 & &0.24\\

		\bottomrule
\end{tabular}}
\end{table}

\subsection{Genome-wide Association Study}
We then applied the copula-based model on the whole genome to search for top gene regions that are associated with the AMD progression. Again, all gene regions were extracted based on GRCh37/hg19 assembly from UCSC Genome Browser and those with less than 10 SNPs were excluded from the analysis. In total we analyzed 22,747 gene regions.

Table \ref{tab:GWAS_top} presents $p$-values of top gene regions identified by the proposed copula-FLM approach, at the $\alpha$ level of $1.0 \times 10^{-5} $. All methods have generated consistent $p$-values for these regions, with copula-FLM producing the smallest $p$-values. The last column presents the number of significant SNPs identified by the traditional single-SNP-based clayton copula model under the nominal $\alpha$ level of 0.05. In addition to the \textit{CFH} and \textit{ARMS2} regions, \textit{DLGAP2-AS1}, \textit{GABRA5}, \textit{CLEC4GP1} and \textit{HAL-C} regions all have more than 30\% of the SNPs to be associated with the AMD progression in a less stringent sense, which explains the reason why they were not among the top genes list in the single-SNP-based GWAS studies but were identified in a gene-based analysis. Among these gene regions, gene \textit{HLA-C} has been shown to have association with AMD in combination with inhibitory Killer-cell immunoglobulin-like receptors on natural killer (NK) cells \citep{HLA_2005}.

\begin{table}
\centering
\caption{Top gene regions identified by GWAS on AREDS data using the copula-FLM approach}
\label{tab:GWAS_top}
\resizebox{\linewidth}{!}{% Resize table to fit within \linewidth horizontally
 \begin{tabular}{llllllc}
 \hline
	Gene & Chr & \# of SNPs & Cop-Score & Cox-Rst  & CoxFLM  & \# of Sig. SNP (\%)*\\	
\hline
	
\textit{CFH} & 1  & 281 & $9.93 \times 10^{-10}$ & $1.43 \times 10^{-7}$ & $4.14 \times 10^{-4}$ & 186 $(66.2)$\\
\textit{PCDH9-AS4} & 13& 55 & $6.64 \times 10^{-8}$ &$4.21 \times 10^{-4}$& $3.85 \times 10^{-3}$ & 9 $(16.4)$\\
\textit{DLGAP2-AS1} &8& 336& $8.40 \times 10^{-7}$& $7.53 \times 10^{-6}$ & $2.14 \times 10^{-5}$ & 229 $(68.2)$\\
\textit{ARMS2} &10 & 51 & $2.07 \times 10^{-6}$ & $1.37 \times 10^{-3}$ & $1.91 \times 10^{-2}$ & 33 $(64.7)$\\
\textit{LINC00476} &9 & 201& $2.23 \times 10^{-6}$ &$3.41 \times 10^{-5}$ & $2.52 \times 10^{-2}$ & 57 $(28.4)$\\
\textit{GABRA5} &15 &168 & $2.59 \times 10^{-6}$ & $7.75 \times 10^{-6}$ & $6.18 \times 10^{-4}$ &68 $(40.5)$\\
\textit{CLEC4GP1} &19&75 & $3.23 \times 10^{-6}$ & $2.60 \times 10^{-4}$ &$4.61 \times 10^{-4}$ &30 $(40.0)$\\
\textit{HLA-C} &6 & 574 & $3.12 \times 10^{-6}$ & $3.85 \times 10^{-4}$ & $8.06 \times 10^{-4}$ &187 $(32.6)$\\
\textit{SULF1} &8 &563 &$8.72 \times 10^{-6}$ &$4.91 \times 10^{-5}$ & $1.40 \times 10^{-3}$ & 75 $(13.3)$\\
\hline
\multicolumn{4}{l}{\small * based on 0.05 alpha level}\\
\end{tabular}}
\end{table}

%\begin{figure}[!ht]
%\centering
%\includegraphics[scale=0.4]{top5_gamma}
%\caption{\label{gamma5}Estimated Effects of Top Genes from GWAS on 6 Bspline Bases using Common and Rare Variants}
%\end{figure}

\section{Conclusion and Discussion}
In this work, we have developed a new gene-based association analysis method for bivariate time-to-event data using the functional linear model under the copula framework. We implemented our proposed methods in R. The key functions can be found in the GitHub https://github.com/yingding99/CopulaFLM. This approach has also been added to a newly developed R package \{CopulaCenR\}. On one hand, the genetic effect can be viewed as a function of the physical positions of variants under the functional linear model. On the other hand, the copula model can effectively handle the correlation between two margins. Combining the FLM with the copula framework fully takes the advantage of both methods for detecting gene regions that are significantly associated with the progression of bilateral diseases.
The great advantage of the proposed copula-FLM model is that the genetic effects are treated as a function of the physical positions of the variants. Therefore, the LD information can be directly taken account in this method.

Extensive simulation studies were performed to evaluate the type-I error rates and power performance of our method. Both the score test and the LRT from our copula FLM model control type-I error well. For the power analysis, our bivariate tests show great advantage by utilizing all available data without collapsing them into the subject level.

We successfully applied our method on AREDS data at four known AMD risk gene regions and obtained consistent findings. Through the genome-wide study, nine gene regions were found to be significantly associated with the disease progression, where \textit{PCDH9-AS4}, \textit{DLGAP2-AS1}, \textit{LINC00476}, \textit{GABRA5}, \textit{CLEC4GP1}, \textit{HLA-C} and \textit{SULF1} are novel regions. The findings from this research provide new perspectives on the genetic underpinning of AMD progression, which will be valuable to establish novel and reliable predictive models for AMD progression. The proposed method is useful for genome-wide association studies of any bilateral disease with survival traits to identify disease susceptible gene regions.

\section*{Acknowledgements}

This research is supported by the National Institute of Health (EY024226). We would like to thank the participants in the AREDS study, who made this research possible, and International AMD Genomics Consortium for generating the genetic data and performing quality check.

%\section*{Supplementary Materials}

%Web Appendix A, referenced in Section~\ref{s:model}, is available with
%this paper at the Biometrics website on Wiley Online
%Library.\vspace*{-8pt}

%  Here, we create the bibliographic entries manually, following the
%  journal style.  If you use this method or use natbib, PLEASE PAY
%  CAREFUL ATTENTION TO THE BIBLIOGRAPHIC STYLE IN A RECENT ISSUE OF
%  THE JOURNAL AND FOLLOW IT!  Failure to follow stylistic conventions
%  just lengthens the time spend copyediting your paper and hence its
%  position in the publication queue should it be accepted.

%  We greatly prefer that you incorporate the references for your
%  article into the body of the article as we have done here
%  (you can use natbib or not as you choose) than use BiBTeX,
%  so that your article is self-contained in one file.
%  If you do use BiBTeX, please use the .bst file that comes with
%  the distribution.  In this case, replace the thebibliography
%  environment below by
%
%  \bibliographystyle{biom}
% \bibliography{mybibilo.bib}

\bibliographystyle{biom}
\bibliography{bibGene}

\begin{thebibliography}{}

\bibitem[\protect\citeauthoryear{{\relax Age-Related Eye Disease Study Research
  Group}}{{\relax Age-Related Eye Disease Study Research Group}}{1999}]{AREDS}
{\relax Age-Related Eye Disease Study Research Group} (1999).
\newblock The age-related eye disease study (areds): design implications. areds
  report no. 1.
\newblock {\em Control Clinical Trials} {\bf 20,} 573--600.

\bibitem[\protect\citeauthoryear{Chen, Lumley, Brody, Heard-Costa, Fox,
  Cupples, and Dupuis}{Chen et~al.}{2014}]{chen_SKAT}
Chen, H., Lumley, T., Brody, J., Heard-Costa, N.~L., Fox, C.~S., Cupples,
  L.~A., and Dupuis, J. (2014).
\newblock Sequence kernel association test for survival traits.
\newblock {\em Genet Epidemiol} {\bf 38,} 191--197.

\bibitem[\protect\citeauthoryear{Chen, Stambolian, Edwards, Branham, Othman,
  and al.}{Chen et~al.}{2010}]{Wei_PNAS_2010}
Chen, W., Stambolian, D., Edwards, A.~O., Branham, K.~E., Othman, M., and al.,
  J.~J. (2010).
\newblock Genetic variants near timp3 and high-density lipoprotein–associated
  loci influence susceptibility to age-related macular degeneration.
\newblock {\em PNAS} {\bf 107,} 7401--7406.

\bibitem[\protect\citeauthoryear{Chien, Bowden, and Chiu}{Chien
  et~al.}{2017}]{burden_family}
Chien, L.-C., Bowden, D., and Chiu, Y.-F. (2017).
\newblock Region-based association tests for sequencing data on survival
  traits.
\newblock {\em Genetic epidemiology} {\bf 41,} 511--522.

\bibitem[\protect\citeauthoryear{Clayton}{Clayton}{1978}]{Clayton}
Clayton, D.~G. (1978).
\newblock A model for association in bivariate life tables and its application
  in epidemiological studies of familial tendency in chronic disease incidence.
\newblock {\em Biometrika} {\bf 65,} 141--151.

\bibitem[\protect\citeauthoryear{de~Boor}{de~Boor}{2011}]{deboor_spline}
de~Boor, C. (2011).
\newblock {\em A Practical Guide to Splines (Applied Mathematical Sciences)}.
\newblock Springer.

\bibitem[\protect\citeauthoryear{Ding, Liu, Yan, Fritsche, Cook, Clemons,
  Ratnapriya, Klein, Abecasis, Swaroop, Chew, Weeks, and Chen}{Ding
  et~al.}{2017}]{prediction_genetics}
Ding, Y., Liu, Y., Yan, Q., Fritsche, L.~G., Cook, R.~J., Clemons, T.,
  Ratnapriya, R., Klein, M.~L., Abecasis, G.~R., Swaroop, A., Chew, E.~Y.,
  Weeks, D.~E., and Chen, W. (2017).
\newblock Bivariate analysis of age-related macular degeneration progression
  using genetic risk scores.
\newblock {\em Genetics} {\bf 206,} 119--133.

\bibitem[\protect\citeauthoryear{Fan, Wang, Mills, Carter, Lobach, Wilson,
  Bailey-Wilson, Weeks, , and Xiong}{Fan et~al.}{2014}]{Ruzong_binary}
Fan, R., Wang, Y., Mills, J.~L., Carter, T.~C., Lobach, I., Wilson, A.~F.,
  Bailey-Wilson, J.~E., Weeks, D.~E., , and Xiong, M. (2014).
\newblock Generalized functional linear models for case-control association
  studies.
\newblock {\em Genet Epidemiol} {\bf 38,} 622--637.

\bibitem[\protect\citeauthoryear{Fan, Wang, Mills, Wilson, Bailey-Wilson, , and
  Xiong}{Fan et~al.}{2013}]{Ruzong_quantative}
Fan, R., Wang, Y., Mills, J.~L., Wilson, A.~F., Bailey-Wilson, J.~E., , and
  Xiong, M. (2013).
\newblock Functional linear models for association analysis of quantitative
  traits.
\newblock {\em Genet Epidemiol} {\bf 37,} 726 -- 742.

\bibitem[\protect\citeauthoryear{Fan, Wang, Yan, Ding, Weeks, Lu, Ren, Cook,
  Xiong, Swaroop, Chew, , and Chen}{Fan et~al.}{2016}]{Ruzong_surv}
Fan, R., Wang, Y., Yan, Q., Ding, Y., Weeks, D., Lu, Z., Ren, H., Cook, R.~J.,
  Xiong, M., Swaroop, A., Chew, E.~Y., , and Chen, W. (2016).
\newblock Gene-based association analysis for censored traits via fixed effect
  functional regressions.
\newblock {\em Genet Epidemiol} {\bf 40,} 133--143.

\bibitem[\protect\citeauthoryear{Fritsche, Chen, and et~al.}{Fritsche
  et~al.}{2013}]{AMD_genetic_2013}
Fritsche, L.~G., Chen, W., and et~al., M.~S. (2013).
\newblock Seven new loci associated with age-related macular degeneration.
\newblock {\em Nature Genetics} {\bf 45,} 433--439.

\bibitem[\protect\citeauthoryear{Fritsche, Igl, Baileyet,
  et~al\mbox{.}}{Fritsche et~al.}{2016}]{AMD_genetic_2016}
Fritsche, L.~G., Igl, W., Baileyet, J. N.~C., et~al. (2016).
\newblock A large genome-wide association study of age-related macular
  degeneration highlights contributions of rare and common variants.
\newblock {\em Nature Genetics} {\bf 48,} 134--143.

\bibitem[\protect\citeauthoryear{Goverdhan, Howell, Mullins, Osmond, Hodgkins,
  Self, Avery, and Lotery}{Goverdhan et~al.}{2005}]{HLA_2005}
Goverdhan, S.~V., Howell, M.~W., Mullins, R.~F., Osmond, C., Hodgkins, P.~R.,
  Self, J., Avery, K., and Lotery, A.~J. (2005).
\newblock Association of hla class i and class ii polymorphisms with
  age-related macular degeneration.
\newblock {\em Investigative Ophthalmology \& Visual Science} {\bf 46,}
  1726--1734.

\bibitem[\protect\citeauthoryear{Gumbel}{Gumbel}{1960}]{Gumbel}
Gumbel, E.~J. (1960).
\newblock Bivariate exponential distributions.
\newblock {\em Journal of the American Statistical Association} {\bf 55,}
  698--707.

\bibitem[\protect\citeauthoryear{Han and Pan}{Han and Pan}{2010}]{burden_test}
Han, F. and Pan, W. (2010).
\newblock A data-adaptive sum test for disease association with multiple common
  or rare variants.
\newblock {\em Human Heredity} {\bf 70,} 537--545.

\bibitem[\protect\citeauthoryear{Hougaard}{Hougaard}{2000}]{Hougaard_2000}
Hougaard, P. (2000).
\newblock {\em Analysis of Multivariate Survival Data}.
\newblock Springer New York.

\bibitem[\protect\citeauthoryear{Joe}{Joe}{1997}]{Joe}
Joe, H. (1997).
\newblock {\em Multivariate models and dependence concepts}.
\newblock Chapman \& Hall.

\bibitem[\protect\citeauthoryear{Li and Leal}{Li and
  Leal}{2008}]{burden_test_Li}
Li, B. and Leal, S.~M. (2008).
\newblock Methods for detecting associations with rare variants for common
  diseases: Application to analysis of sequence data.
\newblock {\em American Journal of Human Genetics} {\bf 83,} 311--321.

\bibitem[\protect\citeauthoryear{Oakes}{Oakes}{1982}]{Oaks}
Oakes, D. (1982).
\newblock A model for association in bivariate survival data.
\newblock {\em Journal of the Royal Statistical Society. Series B} {\bf 44,}
  414--422.

\bibitem[\protect\citeauthoryear{Ramsay, Hooker, and Graves}{Ramsay
  et~al.}{2009}]{FDA_r}
Ramsay, J.~O., Hooker, G., and Graves, S. (2009).
\newblock {\em Functional Data Analysis With R and Matlab}.
\newblock Springer New York.

\bibitem[\protect\citeauthoryear{Sardell, Persad, Pan, and et~al.}{Sardell
  et~al.}{2016}]{Sardell_AMD_CFH}
Sardell, R.~J., Persad, P.~J., Pan, S.~S., and et~al., P.~W. (2016).
\newblock {Progression rate from intermediate to advanced Age-Related Macular
  Degeneration is correlated with the wumber of risk alleles at the CFH locus}.
\newblock {\em Invest Ophthalmol Vis Sci} {\bf 57,} 6107--6115.

\bibitem[\protect\citeauthoryear{Schaffner, Foo, Gabriel1, Reich, Daly1, and
  Altshuler}{Schaffner et~al.}{2005}]{schaffner_LD}
Schaffner, S.~F., Foo, C., Gabriel1, S., Reich, D., Daly1, M.~J., and
  Altshuler, D. (2005).
\newblock Calibrating a coalescent simulation of human genome sequence
  variation.
\newblock {\em Genome Res} {\bf 15,} 1576--1583.

\bibitem[\protect\citeauthoryear{Sklar}{Sklar}{1959}]{sklar}
Sklar, A. (1959).
\newblock Fonctions de repartition a n dimensions et leurs marges.
\newblock {\em Publications de L Institut de Statistique de L Universite de
  Paris} {\bf 8,} 229--231.

\bibitem[\protect\citeauthoryear{Sun, Liu, J.~Cook, Chen, and Ding}{Sun
  et~al.}{2019}]{CopulaRC}
Sun, T., Liu, Y., J.~Cook, R., Chen, W., and Ding, Y. (2019).
\newblock Copula-based score test for bivariate time-to-event data, with
  application to a genetic study of amd progression.
\newblock {\em Lifetime Data Analysis} .

\bibitem[\protect\citeauthoryear{Swaroop, Chew, Abecasis,
  et~al\mbox{.}}{Swaroop et~al.}{2009}]{AMD_prog_1}
Swaroop, A., Chew, E.~Y., Abecasis, G.~R., et~al. (2009).
\newblock Unraveling a multifactorial late-onset disease: from genetic
  susceptibility to disease mechanisms for age-related macular degeneration.
\newblock {\em Annual review of genomics and human genetics} {\bf 10,} 19--43.

\bibitem[\protect\citeauthoryear{{\relax The Eye Diseases Prevalence Research
  Group}}{{\relax The Eye Diseases Prevalence Research Group}}{2004}]{AMD_Prev}
{\relax The Eye Diseases Prevalence Research Group} (2004).
\newblock {Causes and Prevalence of Visual Impairment Among Adults in the
  United States}.
\newblock {\em Archives of Ophthalmology} {\bf 122,} 477--485.

\bibitem[\protect\citeauthoryear{Voorman, Brody, Chen, Lumley, and
  Davis}{Voorman et~al.}{2016}]{seqMeta}
Voorman, A., Brody, J., Chen, H., Lumley, T., and Davis, B. (2016).
\newblock {\em cluster: Meta-Analysis of Region-Based Tests of Rare DNA
  Variants}.

\bibitem[\protect\citeauthoryear{Wei, Lin, and Weissfeld}{Wei
  et~al.}{1989}]{LWL}
Wei, L.~J., Lin, D., and Weissfeld, L. (1989).
\newblock Regression analysis of multivariate incomplete failure time data by
  modeling marginal distributions.
\newblock {\em Journal of the American Statistical Association} {\bf 84,}
  1065--1073.

\bibitem[\protect\citeauthoryear{Wu, Lee, Cai, Li, Boehnke, , and Lin}{Wu
  et~al.}{2011}]{MichealWu_SKAT}
Wu, M.~C., Lee, S., Cai, T., Li, Y., Boehnke, M., , and Lin, X. (2011).
\newblock Rare-variant association testing for sequencing data with the
  sequence kernel association test.
\newblock {\em Am J Hum Genet} {\bf 89,} 82--93.

\bibitem[\protect\citeauthoryear{Wu, Guan, Liu, Novelo, and Bandyopadhyay}{Wu
  et~al.}{2018}]{adaptive-burden}
Wu, X., Guan, T., Liu, D.~J., Novelo, L. G.~L., and Bandyopadhyay, D. (2018).
\newblock Adaptive-weight burden test for associations between quantitative
  traits and genotype data with complex correlations.
\newblock {\em Annals of Applied Statistics} {\bf 12,} 1558--1582.

\bibitem[\protect\citeauthoryear{Yan, Ding, Liu, Sun, Fritsche, Clemons,
  Ratnapriya, Klein, Cook, Liu, Fan, Wei, Abecasis, Swaroop, Chew, Group,
  Weeks, and Chen}{Yan et~al.}{2018}]{Yan2017_AMD}
Yan, Q., Ding, Y., Liu, Y., Sun, T., Fritsche, L.~G., Clemons, T., Ratnapriya,
  R., Klein, M.~L., Cook, R.~J., Liu, Y., Fan, R., Wei, L., Abecasis, G.~R.,
  Swaroop, A., Chew, E.~Y., Group, A.~R., Weeks, D.~E., and Chen, W. (2018).
\newblock Genome-wide analysis of disease progression in age-related macular
  degeneration.
\newblock {\em Human Molecular Genetics} {\bf 27,} 929--940.

\end{thebibliography}

\label{lastpage}

\end{document}